\documentclass[12pt]{article}

\usepackage[a4paper]{geometry}
\usepackage{amssymb, amsmath, amsthm}
\usepackage{enumitem, graphicx, cite, color}
\usepackage[hyphens]{url}
\usepackage[colorlinks=true, citecolor=black, linkcolor=black, urlcolor=blue]{hyperref}

\usepackage{tikz, float, subfig}
\usetikzlibrary{automata}

\theoremstyle{plain}
\newtheorem{theorem}{Theorem}
\newtheorem{lemma}[theorem]{Lemma}
\newtheorem{corollary}[theorem]{Corollary}
\newtheorem{proposition}[theorem]{Proposition}

\theoremstyle{definition}
\newtheorem{definition}[theorem]{Definition}
\newtheorem{problem}[theorem]{Problem}
\newtheorem{question}[theorem]{Question}

\theoremstyle{remark}
\newtheorem*{remark}{Remark}

\newcommand{\doi}[1]{\url{https://doi.org/#1}}

\DeclareMathOperator{\Var}{Var}
\DeclareMathOperator{\Sym}{Sym}

\usepackage{listings}
\definecolor{dkgreen}{RGB}{1,100,32}
\definecolor{mauve}{rgb}{0.58,0,0.82}

\lstdefinestyle{cplusplus} {
    language = C++,
    basicstyle = footnotesizettfamily,
    showspaces = false,
    showstringspaces = false,
    breakautoindent = true,
    flexiblecolumns = true,
    keepspaces = true,
    stepnumber = 1,
    xleftmargin = 0pt
}
\def\dj{d\kern-0.4em\char"16\kern-0.1em}
\def\Dj{\hbox{\raise0.3ex\hbox{-}\kern-0.4em  D}}

\usepackage{authblk}

\title{Counting the number of inequivalent\\ arithmetic expressions on $n$ variables}

\author[1]{Ivan Stošić}
\author[2,3,4]{Ivan Damnjanović}
\author[5]{Žarko Ranđelović}
\affil[1]{Faculty of Sciences and Mathematics, University of Niš, Niš, Serbia}
\affil[2]{Faculty of Mathematics, Natural Sciences and Information Technologies, University of Primorska, Koper, Slovenia}
\affil[3]{Faculty of Electronic Engineering, University of Niš, Niš, Serbia}
\affil[4]{Diffine LLC, San Diego, California, USA}
\affil[5]{Centre for Mathematical Sciences, University of Cambridge, Cambridge, UK}

\date{}

\begin{document}

\maketitle

\begin{abstract}
An expression is any mathematical formula that contains certain formal variables and operations to be executed in a specified order. In computer science, it is usually convenient to represent each expression in the form of an expression tree. Here, we consider only arithmetic expressions, i.e., those that contain only the four standard arithmetic operations:\ addition, subtraction, multiplication and division, alongside additive inversion. We first provide certain theoretical results concerning the equivalence of such expressions and then disclose a $\Theta(n^2)$ algorithm that computes the number of inequivalent arithmetic expressions on $n$ distinct variables.
\end{abstract}


\section{Introduction}

Every mathematical formula is structured and interpreted according to commonly known rules, which define the order in which the mathematical symbols are written and what their meaning and scope are. Here, we constrain ourselves to arithmetic expressions, i.e., mathematical formulae that only contain the four binary arithmetic operations:\ addition, subtraction, multiplication and division, together with additive inversion. We will explore the general form of these expressions, where the operands are formal variables.

In standard computer science literature, an arithmetic expression can be represented by an ordered rooted tree known as the expression tree (see, for example, \cite{gopal2010, grune2012, preiss1999}). The leaves of this tree are operands, while its internal nodes correspond to unary or binary operations. A node representing additive inversion must have exactly one child, while any node corresponding to the four standard arithmetic operations necessarily has exactly two children. Note that in case of noncommutative operations such as subtraction or division, the order of the node's children matters. The value of an expression tree can be computed recursively by applying the operation from the corresponding internal node to the values obtained from the subtrees of its children nodes.

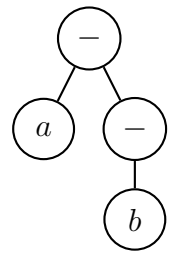
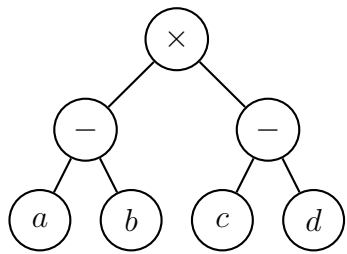
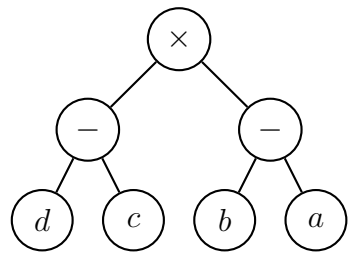
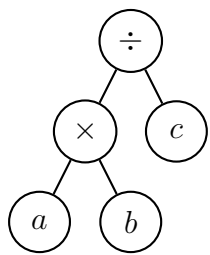
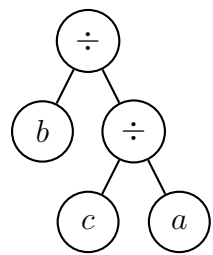
\begin{figure}[H]
    \centering
    \subfloat[The expression $a + b$.] {
        \begin{tikzpicture}[scale=0.6]
            \node[state, minimum size=0.8cm, thick, white] (6) at (-4, -4) {};
            \node[state, minimum size=0.8cm, thick, white] (7) at (4, -4) {};
        
            \node[state, minimum size=0.8cm, thick] (1) at (0, 0) {$+$};
            \node[state, minimum size=0.8cm, thick] (2) at (-1, -2) {$a$};
            \node[state, minimum size=0.8cm, thick] (3) at (1, -2) {$b$};
            
            \draw[thick] (1) to (2);
            \draw[thick] (1) to (3);
        \end{tikzpicture}
        \label{expression_trees_a}
    }
    \hspace{0.5cm}
    \subfloat[The expression $a - (-b)$.] {
        \begin{tikzpicture}[scale=0.6]
            \node[state, minimum size=0.8cm, thick, white] (6) at (-4, -4) {};
            \node[state, minimum size=0.8cm, thick, white] (7) at (4, -4) {};
        
            \node[state, minimum size=0.8cm, thick] (1) at (0, 0) {$-$};
            \node[state, minimum size=0.8cm, thick] (2) at (-1, -2) {$a$};
            \node[state, minimum size=0.8cm, thick] (3) at (1, -2) {$-$};
            \node[state, minimum size=0.8cm, thick] (4) at (1, -4) {$b$};
            
            \draw[thick] (1) to (2);
            \draw[thick] (1) to (3);
            \draw[thick] (3) to (4);
        \end{tikzpicture}
        \label{expression_trees_b}
    }
    \\
    \subfloat[The expression $(a-b) \times (c-d)$.] {
        \begin{tikzpicture}[scale=0.6]
            \node[state, minimum size=0.8cm, thick, white] (6) at (-4, -4) {};
            \node[state, minimum size=0.8cm, thick, white] (7) at (4, -4) {};
        
            \node[state, minimum size=0.8cm, thick] (1) at (0, 0) {$\times$};
            \node[state, minimum size=0.8cm, thick] (2) at (-2, -2) {$-$};
            \node[state, minimum size=0.8cm, thick] (3) at (2, -2) {$-$};
            \node[state, minimum size=0.8cm, thick] (4) at (-3, -4) {$a$};
            \node[state, minimum size=0.8cm, thick] (5) at (-1, -4) {$b$};
            \node[state, minimum size=0.8cm, thick] (6) at (1, -4) {$c$};
            \node[state, minimum size=0.8cm, thick] (7) at (3, -4) {$d$};
            
            \draw[thick] (1) to (2);
            \draw[thick] (1) to (3);
            \draw[thick] (2) to (4);
            \draw[thick] (2) to (5);
            \draw[thick] (3) to (6);
            \draw[thick] (3) to (7);
        \end{tikzpicture}
        \label{expression_trees_c}
    }
    \hspace{0.5cm}
    \subfloat[The expression $(d-c) \times (b-a)$.] {
        \begin{tikzpicture}[scale=0.6]
            \node[state, minimum size=0.8cm, thick, white] (6) at (-4, -4) {};
            \node[state, minimum size=0.8cm, thick, white] (7) at (4, -4) {};
        
            \node[state, minimum size=0.8cm, thick] (1) at (0, 0) {$\times$};
            \node[state, minimum size=0.8cm, thick] (2) at (-2, -2) {$-$};
            \node[state, minimum size=0.8cm, thick] (3) at (2, -2) {$-$};
            \node[state, minimum size=0.8cm, thick] (4) at (-3, -4) {$d$};
            \node[state, minimum size=0.8cm, thick] (5) at (-1, -4) {$c$};
            \node[state, minimum size=0.8cm, thick] (6) at (1, -4) {$b$};
            \node[state, minimum size=0.8cm, thick] (7) at (3, -4) {$a$};
            
            \draw[thick] (1) to (2);
            \draw[thick] (1) to (3);
            \draw[thick] (2) to (4);
            \draw[thick] (2) to (5);
            \draw[thick] (3) to (6);
            \draw[thick] (3) to (7);
        \end{tikzpicture}
        \label{expression_trees_d}
    }
    \\
    \subfloat[The expression $(a \times b) \div c$.] {
        \begin{tikzpicture}[scale=0.6]
            \node[state, minimum size=0.8cm, thick, white] (6) at (-4, -4) {};
            \node[state, minimum size=0.8cm, thick, white] (7) at (4, -4) {};
        
            \node[state, minimum size=0.8cm, thick] (1) at (0, 0) {$\div$};
            \node[state, minimum size=0.8cm, thick] (2) at (-1, -2) {$\times$};
            \node[state, minimum size=0.8cm, thick] (3) at (1, -2) {$c$};
            \node[state, minimum size=0.8cm, thick] (4) at (-2, -4) {$a$};
            \node[state, minimum size=0.8cm, thick] (5) at (0, -4) {$b$};
            
            \draw[thick] (1) to (2);
            \draw[thick] (1) to (3);
            \draw[thick] (2) to (4);
            \draw[thick] (2) to (5);
        \end{tikzpicture}
        \label{expression_trees_e}
    }
    \hspace{0.5cm}
    \subfloat[The expression $b \div (c \div a)$.] {
        \begin{tikzpicture}[scale=0.6]
            \node[state, minimum size=0.8cm, thick, white] (6) at (-4, -4) {};
            \node[state, minimum size=0.8cm, thick, white] (7) at (4, -4) {};
        
            \node[state, minimum size=0.8cm, thick] (1) at (0, 0) {$\div$};
            \node[state, minimum size=0.8cm, thick] (2) at (-1, -2) {$b$};
            \node[state, minimum size=0.8cm, thick] (3) at (1, -2) {$\div$};
            \node[state, minimum size=0.8cm, thick] (4) at (0, -4) {$c$};
            \node[state, minimum size=0.8cm, thick] (5) at (2, -4) {$a$};
            
            \draw[thick] (1) to (2);
            \draw[thick] (1) to (3);
            \draw[thick] (3) to (4);
            \draw[thick] (3) to (5);
        \end{tikzpicture}
        \label{expression_trees_f}
    }
    \caption{The expression tree representations of various arithmetic expressions containing the standard four arithmetic operations, as well as additive inversion.}
\end{figure}

We define two arithmetic expressions to be equivalent if they correspond to the same formula, i.e., if one of their formulae can be obtained from the other by applying standard rules of algebra. For example, it is clear that $a + b = a - (-b)$, which directly implies that the expressions obtained from the trees given in Figures~\ref{expression_trees_a} and \ref{expression_trees_b} are equivalent. Furthermore, the expressions representable via the trees depicted in Figures \ref{expression_trees_c} and \ref{expression_trees_d} are also equivalent. Here, we note that although the functions $(a, b, c) \mapsto (a \times b) \div c$ and $(a, b, c) \mapsto b \div (c \div a)$ do not formally have the same domain, we will consider the expressions shown in Figures~\ref{expression_trees_e} and \ref{expression_trees_f} to be equivalent nonetheless. The reason why this is so is because we are interested only in the obtained formal expressions without regarding them as functions.

It is natural to ask how many inequivalent arithmetic expressions there are under certain constraints. In fact, the problem of finding the number of inequivalent arithmetic expressions on $n$ distinct variables that contain the four standard arithmetic operations:\ addition, subtraction, multiplication and division, was considered by Du \cite{du2008}. Radcliffe \cite{radcliffe2012} also considered the similar problem of finding the number of inequivalent arithmetic expressions on $n$ distinct variables where the same four operations are allowed together with additive inversion. Some results concerning these two problems were obtained in \cite{zhang2018_a}, where equations regarding the exponential generating functions were found. Additionaly, the asymptotic behavior of the two corresponding sequences was analyzed in \cite{zhang2018_b}.

The main result of this paper is a simple and efficient algorithm that computes the number of inequivalent arithmetic expressions on $n$ distinct variables containing the standard four arithmetic operations together with additive inversion. The algorithm uses $\Theta(n^2)$ arbitrary precision integer arithmetic operations and $\Theta(n)$ memory, measured in the number of stored arbitrary precision integers.

The paper shall be structured as follows. Section \ref{sc_prel} will contain the mathematical theory required to formally define the problem in terms of polynomial theory. It will also introduce the auxiliary concepts of sum-type and product-type expressions. Afterwards, Section \ref{sc_expressions} will serve to prove certain mathematical properties concerning the equivalence of sum-type and product-type expressions. Section \ref{sc_algorithm} will provide the overview of the problem's algorithmic solution, whose implementation is given in Appendix \ref{sc_code}. Finally, we disclose some open problems in Section~\ref{sc_conclusion}.

\section{Preliminaries}\label{sc_prel}

In the present section, we will preview the mathemathical theory needed to properly follow the rest of the paper, with a particular focus on polynomial theory. Besides that, we will introduce certain auxiliary terms and concepts that will play a big role while proving the validity of the algorithm disclosed in Section \ref{sc_algorithm}.

To begin, let $X = \{x_1, x_2, x_3, \ldots \}$ be a countably infinite set consisting of formal variables and let $\mathbb{Z}[X]$ be the ring of polynomials in these variables with integer coefficients. It is well known that $\mathbb{Z}[X]$ represents an integral domain with a unique factorization, i.e., a unique factorization domain (see, for example, \cite{lang2002}). Furthermore, let $\mathbb{Z}(X)$ denote the field of fractions of $\mathbb{Z}[X]$. For each fraction $f \in \mathbb{Z}(X)$, we will say that $f = \frac{P}{Q}$ is a \emph{canonical representation} of $f$ provided $P, Q \in \mathbb{Z}[X]$ are two coprime polynomials such that $Q \not\equiv 0$. It is straightforward to see that each $f \in \mathbb{Z}(X)$ has precisely two distinct canonical representations $f = \frac{P_1}{Q_1}$ and $f = \frac{P_2}{Q_2}$ that are related by the equalities $P_2 = -P_1$ and $Q_2 = -Q_1$. This immediately follows from the fact that the only two invertible elements in $\mathbb{Z}[X]$ are $1$ and $-1$, hence the only two associates of any nonzero polynomial $P \in \mathbb{Z}[X]$ are $P$ and $-P$. For further insight into the standard polynomial theory, please refer to \cite{dufo2004, lang2002, pinter2009}.

Moving on, let $\mathcal{M} \subset \mathbb{N}_0^\mathbb{N}$ consist of all the $\mathbb{N}$-indexed sequences of nonnegative integers that contain finitely many nonzero elements. Each polynomial $P \in \mathbb{Z}[X]$ can now conveniently be represented by a mapping $\Omega_P \colon \mathcal{M} \to \mathbb{Z}$ that assigns each $M = (m_1, m_2, m_3, \ldots) \in \mathcal{M}$ the coefficient corresponding to the $\prod_{j \in \mathbb{N}} x_j^{m_j}$ term in $P$. Moreover, let $\mathcal{M}_P \subset \mathcal{M}$ denote the set comprising the finitely many $M \in \mathcal{M}$ for which $\Omega_P(M) \neq 0$. For example, if we have $P = 2 x_1 x_2^3 - 4 x_3 x_4 x_6^2$, then
\[
    \Omega_P(1, 3, 0, 0, \ldots) = 2 \qquad \mbox{and} \qquad \Omega_P(0, 0, 1, 1, 0, 2, 0, 0, \ldots) = -4,
\]
while $\Omega_P(M) = 0$ for any other $M \in \mathcal{M}$. Also, it is trivial to obtain $\mathcal{M}_P = \{ (1, 3, 0, 0, \ldots), (0, 0, 1, 1, 0, 2, 0, 0, \ldots) \}$.

We will say that a polynomial $P \in \mathbb{Z}[X]$ \emph{contains} the variable $x_i$ if there exists an $M = (m_1, m_2, \ldots) \in \mathcal{M}_P$ such that $m_i \neq 0$. Thus, a constant polynomial contains no variable. It is now convenient to generalize the aforementioned definition to fractions by saying that $f \in \mathbb{Z}(X)$ \emph{contains} the variable $x_i$ if at least one of the two polynomials $P, Q$ from its canonical representation $f = \frac{P}{Q}$ contains the variable $x_i$. Furthermore, let $\Var(f) \subset X$ signify the finite set of all the variables contained by $f \in \mathbb{Z}(X)$. Here, we observe that $\Var(f) = \Var(P) \cup \Var(Q)$ holds for any canonical representation $f = \frac{P}{Q}$. It is also worth noting that it does not matter which of the two possible canonical representations is being used to find $\Var(f)$, since $\Var(-P)=\Var(P)$ and $\Var(-Q) = \Var(Q)$, hence the concept is well-defined. We resume by disclosing the following auxiliary lemma.

\begin{lemma}\label{var_sum_lemma}
    For a given $s \in \mathbb{N}$, if $f_1, f_2, \ldots, f_s \in \mathbb{Z}(X) \setminus \{ 0 \}$ are fractions such that the sets $\Var(f_1), \Var(f_2), \ldots, \Var(f_s)$ are mutually disjoint, then we have
    \[
        \Var\left( \sum_{i = 1}^s f_i \right) = \bigcup_{i = 1}^s \Var(f_i) .
    \]
\end{lemma}
\begin{proof}
    Let $f_i = \frac{P_i}{Q_i}$ be the canonical representation of $f_i$, for each $1 \le i \le s$, and let $f = \sum_{i = 1}^s f_i$. From here, it immediately follows that $f = \frac{P}{Q}$, where
    \begin{align}
        \label{prel_aux1} P &= \sum_{i = 1}^s \left( P_i \, \prod_{j = 1, \, j\neq i}^s Q_j\right),\\
        \nonumber Q &= \prod_{i = 1}^s Q_i .
    \end{align}
    
    Furthermore, $Q_i$ and $Q_j$ necessarily have a constant GCD for any $1 \le i, j \le s, \linebreak i \neq j$, since otherwise, they would both contain the same variable, hence $\Var(f_i) \cap \Var(f_j) \neq \varnothing$. It is also trivial to conclude that $Q_i$ and $P_j$ must have a constant GCD for any $1 \le i, j \le s$. If we let $R \in \mathbb{Z}[X]$ be a nonconstant irreducible factor of $Q$, this means that $R \mid Q_k$ surely holds for a fixed $1 \le k \le s$, hence $R \nmid P_k$, while $R \nmid P_i, Q_i$ is also necessarily satisfied for every $1 \le i \le s, \, i \neq k$. Therefore, the sum term from Equation~\eqref{prel_aux1} is divisible by $R$ if and only if $i \neq k$, which directly yields $R \nmid P$. Thus, the canonical representation of fraction $f$\linebreak bears the form $f = \frac{P_0}{Q_0}$, where $P_0 = \frac{P}{c}$ and $Q_0 = \frac{Q}{c}$ for some constant $c \in \mathbb{Z}$.

    Bearing in mind that $\Var(f) \subseteq \bigcup_{i = 1}^s \Var(f_i)$ is obviously true, it is sufficient to demonstrate that $\Var(f_i) \subseteq \Var(f)$ holds for each $1 \le i \le s$ in order to complete the proof. Let $x_h \in \Var(f_\ell)$. If $x_h \in \Var(Q_\ell)$, then clearly $x_h \in \Var(Q)$, hence $x_h \in \Var(f)$ and there is nothing left to discuss. On the other hand, if $x_h \not\in \Var(Q_\ell)$, then surely $x_h \in \Var(P_\ell)$, which means that the sum term from Equation~\eqref{prel_aux1} contains a term with a positive power of $x_h$ corresponding to a nonzero coefficient if and only if $i = \ell$. From here, it is easy to see that $x_h \in \Var(P)$, hence $x_h \in \Var(f)$ holds once again, thus yielding $\Var(f_\ell) \subseteq \Var(f)$.
\end{proof}

In an entirely analogous manner, it is possible to obtain the next lemma whose proof we choose to omit.

\begin{lemma}\label{var_prod_lemma}
    For a given $s \in \mathbb{N}, \, t \in \mathbb{N}_0$, if $f_1, f_2, \ldots, f_s, g_1, g_2, \ldots, g_t \in \mathbb{Z}(X) \setminus \{ 0 \}$ are fractions such that the sets $\Var(f_1), \ldots, \Var(f_s), \Var(g_1), \ldots, \Var(g_t)$ are mutually disjoint, then we have
    \[
        \Var\left( \frac{\prod_{i = 1}^s f_i}{\prod_{i = 1}^t g_i} \right) = \bigcup_{i = 1}^s \Var(f_i) \cup \bigcup_{i = 1}^t \Var(g_i).
    \]
\end{lemma}

We will now introduce the auxiliary concepts of a sum-type and a product-type expression. For starters, we define both the \emph{sum-type expression of depth zero} and the \emph{product-type expression of depth zero} to be any $\mathbb{Z}(X)$ element of the form $x_i$ or $-x_i$ for some $i \in \mathbb{N}$. The sum-type and product-type expressions of positive depth can now recursively be defined as follows.

\begin{definition}
    For each $k \in \mathbb{N}$, a \emph{sum-type expression of depth $k \in \mathbb{N}$} is a fraction $f \in \mathbb{Z}(X)$ that can be written as $f = \sum_{i = 1}^s f_i$, where $s \in \mathbb{N}, \, s \ge 2$, and the fractions $f_1, f_2, \ldots, f_s \not\equiv 0$ are such that:
    \begin{enumerate}[label=\textbf{(\roman*)}]
        \item all of them represent a product-type expression of depth at most $k-1$, while at least one of them is a product-type expression of depth exactly $k-1$;
        \item the sets $\Var(f_1), \Var(f_2), \ldots, \Var(f_s)$ are mutually disjoint.
    \end{enumerate}
\end{definition}

\begin{definition}
    For any $k \in \mathbb{N}$, a \emph{product-type expression of depth $k \in \mathbb{N}$} is a fraction $f \in \mathbb{Z}(X)$ that can be written as $f = \dfrac{\prod_{i = 1}^s f_i}{\prod_{i = 1}^t g_i}$, where $s \in \mathbb{N}, \, t \in \mathbb{N}_0, \, s + t \ge 2$, and the fractions $f_1, f_2, \ldots, f_s, g_1, g_2, \ldots, g_t \not\equiv 0$ are such that:
    \begin{enumerate}[label=\textbf{(\roman*)}]
        \item all of them represent a sum-type expression of depth at most $k-1$, while at least one of them is a sum-type expression of depth exactly $k-1$;
        \item the sets $\Var(f_1), \ldots, \Var(f_s), \Var(g_1), \ldots, \Var(g_t)$ are mutually disjoint.
    \end{enumerate}
\end{definition}

Finally, we will say that a \emph{sum-type expression} is a sum-type expression of any depth $k \in \mathbb{N}_0$, while a \emph{product-type expression} is a product-type expression of any depth $k \in \mathbb{N}_0$. We further note that Lemmas~\ref{var_sum_lemma} and \ref{var_prod_lemma} swiftly imply that no sum-type or product-type expression can be zero.

While examining any arithmetic expression, it is possible to recursively transform each subtraction into addition by simply ``forwarding'' the minus into the corresponding operand in the form of a unary minus. Thus, it suffices to consider only the arithmetic expressions whose allowed operations are addition, additive inversion, multiplication and division. Bearing in mind the newly defined concepts of a sum-type and product-type expression together with the results obtained in Lemmas~\ref{var_sum_lemma} and \ref{var_prod_lemma}, it is not difficult to observe that our problem is actually equivalent to the following one.

\begin{problem}\label{problem_reworked}
    For each $n \in \mathbb{N}$, determine how many distinct fractions $f \in \mathbb{Z}(X)$ represent a sum-type or product-type expression such that $\Var(f) = \{x_1, x_2, \ldots, x_n\}$.
\end{problem}

In the upcoming sections, our focus will be to expand the knowledge regarding sum-type and product-type expressions and then to derive a direct solution to Problem~\ref{problem_reworked}.

\section{Sum-type and product-type expressions}\label{sc_expressions}

By virtue of Lemmas~\ref{var_sum_lemma} and \ref{var_prod_lemma}, we may conclude that any sum-type or product-type expression of positive depth surely contains at least two different variables. For this reason, it is impossible for a sum-type or product-type expression of zero depth to be equal to a sum-type or product-type expression of positive depth. It is in fact true that a sum-type expression of positive depth and a product-type expression of positive depth can also never be equal, which leads us to the next theorem.

\begin{theorem}\label{sum_prod_th}
    A sum-type expression of positive depth and a product-type expression of positive depth can never be equal.
\end{theorem}

As it turns out, two sum-type or product-type expressions of positive depth are equal under very specific conditions, as stated in the following two theorems.

\begin{theorem}\label{sum_sum_th}
    Let $\sum_{i=1}^{s_1} f_{1, i}$ and $\sum_{i=1}^{s_2} f_{2, i}$ be two sum-type expressions of positive depth, where $s_1, s_2 \ge 2$ and $f_{1, 1}, f_{1, 2}, \ldots, f_{1, s_1}, f_{2, 1}, f_{2, 2}, \ldots, f_{2, s_2} \in \mathbb{Z}(X)$ are product-type expressions such that:
    \begin{enumerate}[label=\textbf{(\roman*)}]
        \item $\Var(f_{1, 1}), \Var(f_{1, 2}), \ldots, \Var(f_{1, s_1})$ are mutually disjoint;
        \item $\Var(f_{2, 1}), \Var(f_{2, 2}), \ldots, \Var(f_{2, s_2})$ are mutually disjoint.
    \end{enumerate}
    Then $\sum_{i=1}^{s_1} f_{1, i} = \sum_{i=1}^{s_2} f_{2, i}$ holds if and only if $s_2 = s_1$ and there exists a permutation $\pi \in \Sym(s_1)$ such that $f_{2, \pi(i)} = f_{1, i}$ for every $1 \le i \le s_1$.
\end{theorem}

\begin{theorem}\label{prod_prod_th}
    Let $\dfrac{\prod_{i=1}^{s_1} f_{1, i}}{\prod_{i=1}^{t_1} g_{1, i}}$ and $\dfrac{\prod_{i=1}^{s_2} f_{2, i}}{\prod_{i=1}^{t_2} g_{2, i}}$ be two product-type expressions of positive depth, where $s_i \ge 1, t_i \ge 0, s_i + t_i \ge 2$, for $i \in \{1, 2\}$, while $f_{1, 1}, \ldots, f_{1, s_1}, g_{1, 1}, \linebreak \ldots, g_{1, t_1}, f_{2, 1}, \ldots, f_{2, s_2}, g_{2, 1}, \ldots, g_{2, t_2} \in \mathbb{Z}(X)$ are sum-type expressions such that:
    \begin{enumerate}[label=\textbf{(\roman*)}]
        \item $\Var(f_{1, 1}), \ldots, \Var(f_{1, s_1}), \Var(g_{1, 1}), \ldots, \Var(g_{1, t_1})$ are mutually disjoint;
        \item $\Var(f_{2, 1}), \ldots, \Var(f_{2, s_2}), \Var(g_{2, 1}), \ldots, \Var(g_{2, t_2})$ are mutually disjoint.
    \end{enumerate}
    Then $\dfrac{\prod_{i=1}^{s_1} f_{1, i}}{\prod_{i=1}^{t_1} g_{1, i}}$ and $\dfrac{\prod_{i=1}^{s_2} f_{2, i}}{\prod_{i=1}^{t_2} g_{2, i}}$ are equal up to sign if and only if $s_2 = s_1, t_2 = t_1$, and there exist permutations $\pi \in \Sym(s_1)$ and $\sigma \in \Sym(t_1)$ such that $f_{2, \pi(i)}$ and $f_{1, i}$ are equal up to sign for every $1 \le i \le s_1$, while $g_{2, \sigma(j)}$ and $g_{1, j}$ are equal up to sign for every $1 \le j \le t_1$.
\end{theorem}

The goal of the present section will be to provide the full proof of Theorems~\ref{sum_prod_th}, \ref{sum_sum_th} and \ref{prod_prod_th}. These theorems will serve as the backbone to the algorithm disclosed in Section \ref{sc_algorithm} as they directly guarantee the validity behind its logical reasoning. All three of their proofs will heavily rely on formal partial derivatives of $\mathbb{Z}(X)$ fractions. Here, for convenience, we will regard each partial derivative $\frac{\partial f}{\partial x_i}$ as a purely formal concept, i.e., we will always treat it as a $\mathbb{Z}(X)$ element and never as a function. While doing so, it is trivial to verify that all the standard rules of derivative calculus continue to hold. It is also fairly easy to notice that for every $f \in \mathbb{Z}(X)$ and $x_i \in X$, we have $\frac{\partial f}{\partial x_i} \equiv 0$ if and only if $x_i \not\in \Var(f)$. Therefore, the partial derivative of any sum-type or product-type expression with respect to each variable it contains can never be zero. With this in mind, we derive the proof of Theorem \ref{sum_prod_th} as follows.

\bigskip\noindent
\emph{Proof of Theorem \ref{sum_prod_th}}.\quad
Suppose that there exists a sum-type expression of positive depth and a product-type expression of positive depth that are equal. In other words, let $h = \sum_{i = 1}^r h_i = \dfrac{\prod_{i = 1}^s f_i}{\prod_{i = 1}^t g_i}$ hold for some $r \ge 2$ and $s \ge 1, t \ge 0, s + t \ge 2$, where $h_1, h_2, \ldots, h_r$ are product-type expressions with mutually disjoint sets of contained variables and $f_1, f_2, \ldots, f_s, g_1, g_2, \ldots, g_t$ are sum-type expressions also with mutually disjoint sets of contained variables. For convenience, let $f_{s+1} = \frac{1}{g_1}, \linebreak f_{s+2} = \frac{1}{g_2}, \ldots, f_{s+t} = \frac{1}{g_t}$, so that $h = \sum_{i = 1}^r h_i = \prod_{i = 1}^{s + t} f_i$.

We will now prove that any two variables of fraction $h$ contained by different operands from $h_1, h_2, \ldots, h_r$ are surely contained by the same operand from $f_1, f_2, \ldots, f_{s+t}$. Let $x_{\ell_1} \in \Var(h_{k_1})$ and $x_{\ell_2} \in \Var(h_{k_2})$, where $1 \le k_1, k_2 \le r, \linebreak k_1 \neq k_2$. In this case, it is obvious that $\frac{\partial h}{\partial x_{\ell_1}} = \frac{\partial h_{k_1}}{\partial x_{\ell_1}}$, which quickly gives us $x_{\ell_2} \not\in \Var\left(\frac{\partial h}{\partial x_{\ell_1}} \right)$. Now, suppose that $x_{\ell_1} \in \Var(f_{k_3})$ holds for some $1 \le k_3 \le s + t$. Since $x_{\ell_1} \not\in \Var(f_i)$ for every $1 \le i \le s + t, \, i \neq k_3$, we immediately obtain
\begin{equation}\label{maux_1}
    \frac{\partial h}{\partial x_{\ell_1}} = \frac{\partial f_{k_3}}{\partial x_{\ell_1}} \, \prod_{i = 1, \, i \neq k_3}^{s + t} f_i .
\end{equation}
Furthermore, if we have $x_{\ell_2} \not\in \Var(f_{k_3})$, then Equation~\eqref{maux_1} yields $x_{\ell_2} \in \Var\left( \frac{\partial h}{\partial x_{\ell_1}}\right)$, which is impossible. Therefore, $x_{\ell_2} \in \Var(f_{k_3})$ must be true.

Thus, we may conclude that if any two variables $x_{\ell_1}, x_{\ell_2} \in \Var(h)$ are not contained by the same operand from $h_1, \ldots, h_r$, then they are certainly contained by the same operand from $f_1, \ldots, f_{s+t}$. However, if we have two variables $x_{\ell_1}, x_{\ell_2} \in \Var(h)$ that are indeed contained by the same operand from $h_1, \ldots, h_r$, then it is obviously possible to select a distinct third variable $x_{\ell_3}$ that is contained by a different $h_i$-operand, bearing in mind that $r \ge 2$. In this case, it is easy to see that $x_{\ell_1}$ and $x_{\ell_3}$ are necessarily contained by the same $f_i$-operand, as are $x_{\ell_2}$ and $x_{\ell_3}$. Therefore, $x_{\ell_1}$ and $x_{\ell_2}$ are once again contained by the same operand from $f_1, \ldots, f_{s+t}$. This precisely means that there exists an $f_k, \, 1 \le k \le s + t$, such that $\Var(f_k) = \Var(h)$, which is clearly not possible since $s + t \ge 2$. We conclude that a sum-type expression of positive depth and a product-type expression of positive depth cannot be equal. \hfill\qed

\bigskip
Before disclosing the proofs of Theorems \ref{sum_sum_th} and \ref{prod_prod_th}, we will need several additional results. For this reason, we shall first prove the following lemma alongside two of its immediate corollaries.

\begin{lemma}\label{expression_format_lemma}
    The canonical representation $f = \frac{P}{Q}$ of every sum-type or product-type expression $f \in \mathbb{Z}(X)$ is such that:
    \begin{enumerate}[label=\textbf{(\roman*)}]
        \item $\mathcal{M}_P \cap \mathcal{M}_Q = \varnothing$;
        \item for each $M \in \mathcal{M}$, we have $\Omega_P(M), \Omega_Q(M) \in \{-1, 0, 1\}$.
    \end{enumerate}
\end{lemma}
\begin{proof}
    We will carry out the proof by mathematical induction. To begin, it is obvious that the given statements are true for any sum-type or product-type expression of zero depth. Now, suppose that the statements are satisfied for every sum-type or product-type expression of depth less than $k \in \mathbb{N}$.

    If $f \in \mathbb{Z}(X)$ is a sum-type expression of depth $k$, then it can certainly be written as $f = \sum_{i = 1}^s f_i$, where $s \ge 2$ and the fractions $f_1, f_2, \ldots, f_s$ are product-type expressions of depth below $k$ with mutually disjoint sets of contained variables. Furthermore, let $f_i = \frac{P_i}{Q_i}$ be the canonical representation of $f_i$, for each $1 \le i \le s$. From here, we immediately obtain $f = \frac{P}{Q}$, where
    \[
        P = \sum_{i = 1}^s \left( P_i \, \prod_{j = 1, \, j\neq i}^s Q_j\right) \qquad \mbox{and} \qquad Q = \prod_{i = 1}^s Q_i .
    \]
    As already shown in the proof of Lemma \ref{var_sum_lemma}, the polynomials $P$ and $Q$ necessarily have a constant GCD. Moreover, by statement \textbf{(i)} from the induction hypothesis, it is not difficult to observe that among the polynomials $Q$ and $P_i \, \prod_{j = 1, \, j\neq i}^s Q_j, \linebreak 1 \le i \le s$, no two can both contain the same monomial, i.e., two monomials with the same powers of all the variables. Therefore, it promptly follows that $\mathcal{M}_P \cap \mathcal{M}_Q = \varnothing$. It is also trivial to see that $\Omega_P(M), \Omega_Q(M) \in \{-1, 0, 1\}$ must be satisfied for any $M \in \mathcal{M}$. Since $\gcd(P, Q)$ is constant, we further obtain that these two polynomials are certainly coprime, hence $f = \frac{P}{Q}$ represents the canonical representation of $f$. Thus, both \textbf{(i)} and \textbf{(ii)} indeed hold for fraction $f$.

    On the other hand, if $f \in \mathbb{Z}(X)$ is a product-type expression of depth $k$, then it can surely be written as $f = \dfrac{\prod_{i = 1}^s f_i}{\prod_{i = 1}^t g_i}$, where $s \ge 1, \, t \ge 0, \, s + t \ge 2$, and the fractions $f_1, f_2, \ldots, f_s, g_1, g_2, \ldots, g_t$ are sum-type expressions of depth below $k$ with mutually disjoint sets of contained variables. Now, let
    \[
        f_1 = \frac{P_1}{Q_1}, \, f_2 = \frac{P_2}{Q_2}, \, \ldots, \, f_s = \frac{P_s}{Q_s}, \, g_1 = \frac{Q_{s+1}}{P_{s+1}}, \, g_2 = \frac{Q_{s+2}}{P_{s+2}}, \, \ldots, \, g_t = \frac{Q_{s+t}}{P_{s+t}}
    \]
    be the canonical representations of these sum-type expressions, so that $f = \frac{P}{Q}$, where $P = \prod_{i=1}^{s+t} P_i$ and $Q = {\prod_{i=1}^{s+t} Q_i}$. By using the same strategy from Lemma \ref{var_sum_lemma}, it is easy to see that the polynomials $P$ and $Q$ must have a constant GCD. Furthermore, by simply expanding both products from $P$ and $Q$, we conclude that $\mathcal{M}_P \cap \mathcal{M}_Q = \varnothing$ is true and that $\Omega_P(M), \Omega_Q(M) \in \{-1, 0, 1\}$ holds for every $M \in \mathcal{M}$. From here, it also quickly follows that $P$ and $Q$ are necessarily coprime, thus implying that both provided statements are once again satisfied for fraction $f$.
\end{proof}

\begin{corollary}\label{difference_cor}
    Let $f_1, f_2 \in \mathbb{Z}(X)$ each be a sum-type or product-type expression. If $f_1 - f_2$ is a constant, then $f_1 = f_2$.
\end{corollary}
\begin{proof}
    Suppose that $f_1 - f_2 = \frac{c_1}{c_2}$ for some coprime $c_1, c_2 \in \mathbb{Z}, \, c_2 \neq 0$, and let $f_1 = \frac{P_1}{Q_1}$ and $f_2 = \frac{P_2}{Q_2}$ be the canonical representations of $f_1$ and $f_2$, respectively. Thus, we have
    \[
        c_2 \, P_1 Q_2 - c_2 \, P_2 Q_1 = c_1 \, Q_1 Q_2 .
    \]
    It promptly follows that $Q_1 \mid c_2 \, P_1 Q_2$, which then quickly gives $Q_1 \mid Q_2$, since $Q_1$ and $P_1$ are clearly coprime, while statement \textbf{(ii)} from Lemma \ref{expression_format_lemma} guarantees that $Q_1$ and $c_2$ are also coprime. In an analogous manner, it can be shown that $Q_2 \mid Q_1$ must be true, hence $Q_1$ and $Q_2$ are surely associates. Without loss of generality, we may now assume that $Q_1 = Q_2$, which directly leads us to $c_2 \, P_1 - c_2 \, P_2 = c_1 \, Q_1$. Since $c_2 \neq 0$, we see that if $c_1 \neq 0$, then statement \textbf{(i)} from Lemma \ref{expression_format_lemma} immediately yields a contradiction. Therefore, $\frac{c_1}{c_2} = 0$ is necessarily satisfied.
\end{proof}

\begin{corollary}\label{quotient_cor}
    Let $f_1, f_2 \in \mathbb{Z}(X)$ each be a sum-type or product-type expression. If $\frac{f_1}{f_2}$ is a constant, then $f_1$ and $f_2$ are equal up to sign.
\end{corollary}
\begin{proof}
    Suppose that $\frac{f_1}{f_2} = \frac{c_1}{c_2}$ for some coprime $c_1, c_2 \in \mathbb{Z}, \, c_2 \neq 0$, and let $f_1 = \frac{P_1}{Q_1}$ and $f_2 = \frac{P_2}{Q_2}$ be the canonical representations of $f_1$ and $f_2$, respectively. It is obvious that $c_1 \neq 0$. Moreover, a quick computation directly tells us that $c_2 \, P_1 Q_2 = c_1 \, P_2 Q_1$. Since $P_1$ and $Q_1$ are coprime, as are $P_2$ and $Q_2$, it is trivial to conclude that $Q_1 \mid c_2 \, Q_2$ and $Q_2 \mid c_1 \, Q_1$ are both necessarily satisfied. Moreover, statement \textbf{(ii)} from Lemma~\ref{expression_format_lemma} guarantees that $Q_1$ and $Q_2$ are both surely coprime with any noninvertible nonzero constant, hence $Q_1 \mid Q_2$ and $Q_2 \mid Q_1$ also hold, which means that $Q_1$ and $Q_2$ must be associates. Thus, we may assume without loss of generality that $Q_1 = Q_2$, which promptly leads us to $c_2 \, P_1 = c_1 \, P_2$. Bearing in mind statement \textbf{(ii)} from Lemma \ref{expression_format_lemma} once more, it is clear that $c_1$ and $c_2$ are necessarily equal up to sign, which precisely means that $f_1$ and $f_2$ are equal up to sign as well.
\end{proof}

We will now rely on Corollary \ref{difference_cor} to provide the proof of Theorem~\ref{sum_sum_th} and thus demonstrate that any two sum-type expressions of positive depth are equal if and only if their summands are pairwise equal in some order.

\bigskip\noindent
\emph{Proof of Theorem \ref{sum_sum_th}}.\quad
If $s_2 = s_1$ and the given permutation $\pi \in \Sym(s_1)$ exists, it is trivial to observe that $\sum_{i = 1}^{s_1} f_{1, i} = \sum_{i = 1}^{s_2} f_{2, i}$ must be true. Thus, it suffices to only prove the converse. Suppose that $\sum_{i = 1}^{s_1} f_{1, i} = \sum_{i = 1}^{s_2} f_{2, i}$ does indeed hold for two given sum-type expressions, and let $f \in \mathbb{Z}(X)$ denote both of these sums.

To begin, we will demonstrate that if two variables $x_{\ell_1}, x_{\ell_2} \in \Var(f)$ are contained by the same operand from $f_{1, 1}, f_{1, 2}, \ldots, f_{1, s_1}$, then they are necessarily contained by the same operand from $f_{2, 1}, f_{2, 2}, \ldots, f_{2, s_2}$, as well. We will carry out the proof by contradiction. Without loss of generality, let $x_{\ell_1}, x_{\ell_2} \in \Var(f_{1, 1})$ be such that $x_{\ell_1} \in \Var(f_{2, 1})$ and $x_{\ell_2} \in \Var(f_{2, 2})$. In this scenario, it is clear that $f_{1, 1}$ is surely a product-type expression of positive depth, hence it can be represented as $f_{1, 1} = \prod_{i = 1}^t g_i$, where $t \ge 2$ and $g_1, g_2, \ldots, g_t \in \mathbb{Z}(X)$ have mutually disjoint sets of contained variables, while each of them is a sym-type expression or the reciprocal of a sum-type expression. From here, it follows that $x_{\ell_1} \in \Var(g_k)$ holds for some $1 \le k \le t$. Furthermore, we notice that
\[
    \frac{\partial f}{\partial x_{\ell_1}} = \frac{\partial f_{1, 1}}{\partial x_{\ell_1}} = \frac{\partial g_k}{\partial x_{\ell_1}} \, \prod_{i = 1, \, i \neq k}^t g_i ,
\]
hence if $x_{\ell_2} \not\in \Var(g_k)$, then $x_{\ell_2} \in \Var\left(\frac{\partial f}{\partial x_{\ell_1}}\right)$. However, this is clearly impossible given the fact that $\frac{\partial f}{\partial x_{\ell_1}} = \frac{\partial f_{2,1}}{\partial x_{\ell_1}}$. Thus, $x_{\ell_2} \in \Var(g_k)$ must be true. Now, since $t \ge 2$, there certainly exists a variable $x_{\ell_3} \in \Var(f_{1, 1})$ such that $x_{\ell_3} \not\in \Var(g_k)$. From $\frac{\partial f}{\partial x_{\ell_3}} = \frac{\partial f_{1, 1}}{\partial x_{\ell_3}}$, it is possible to show in an entirely analogous manner that $x_{\ell_1}, x_{\ell_2} \in \Var\left(\frac{\partial f}{\partial x_{\ell_3}} \right)$ is surely satisfied. However, $\frac{\partial f}{\partial x_{\ell_3}}$ cannot simultaneously contain both of these variables because they are not contained by the same operand from $f_{2, 1}, f_{2, 2}, \ldots, f_{2, s_2}$. Therefore, we reach a contradiction once again.

Taking everything into consideration, we swiftly obtain that any two variables $x_{\ell_1}, x_{\ell_2} \in \Var(f)$ are contained by the same operand from $f_{1, 1}, f_{1, 2}, \ldots, f_{1, s_1}$ if and only if they are contained by the same operand from $f_{2, 1}, f_{2, 2}, \ldots, f_{2, s_2}$. This observation promptly implies $s_2 = s_1$ and that there necessarily exists a permutation $\pi \in \Sym(s_1)$ such that $\Var(f_{2, \pi(i)}) = \Var(f_{1, i})$ for every $1 \le i \le s_1$. In order to complete the proof, we will show that $f_{2, \pi(i)} = f_{1, i}$ is always satisfied. Now, for any fixed $1 \le k \le s_1$, it is clear that
\[
    f_{2, \pi(k)} - f_{1, k} = \sum_{i = 1, \, i \neq k}^{s_1} \left( f_{1, i} - f_{2, \pi(i)} \right) ,
\]
which tells us that all the variables contained by $f_{2, \pi(k)} - f_{1, k}$ must belong to the set $\bigcup_{i = 1, \, i \neq k}^{s_1} \Var(f_{1, i})$. However, we also already know that $\Var(f_{2, \pi(k)} - f_{1, k}) \subseteq \Var(f_{1, k})$. Since the sets $\Var(f_{1, k})$ and $\bigcup_{i = 1, \, i \neq k}^{s_1} \Var(f_{1, i})$ are obviously disjoint, it follows that the fraction $f_{2, \pi(k)} - f_{1, k}$ contains no variable, i.e., $f_{2, \pi(k)} - f_{1, k}$ is a constant. By virtue of Corollary \ref{difference_cor}, we may conclude that $f_{2, \pi(k)} = f_{1, k}$. Therefore, $f_{2, \pi(i)} = f_{1, i}$ is certainly true for any $1 \le i \le s_1$. \hfill\qed

\bigskip
The rest of the section will focus on showing that any two product-type expressions are equal up to sign if and only if their numerator and denominator factors are pairwise equal up to sign in some order. We will first prove the asserted claim for products of sum-type expressions, as disclosed in the next lemma.

\begin{lemma}\label{pure_product_lemma}
    For a given $s_1, s_2 \in \mathbb{N}_0$, let $f_{1, 1}, f_{1, 2}, \ldots, f_{1, s_1} \in \mathbb{Z}(X)$ be sum-type expressions such that the sets $\Var(f_{1, 1}), \Var(f_{1, 2}), \ldots, \Var(f_{1, s_1})$ are mutually disjoint, and let $f_{2, 1}, f_{2, 2}, \ldots, f_{2, s_2} \in \mathbb{Z}(X)$ be sum-type expressions such that the sets $\Var(f_{2, 1}), \Var(f_{2, 2}), \ldots, \Var(f_{2, s_2})$ are also mutually disjoint. If $\prod_{i = 1}^{s_1} f_{1, i}$ and $\prod_{i = 1}^{s_2} f_{2, i}$ are equal up to sign, then $s_2 = s_1$ and there exists a permutation $\pi \in \Sym(s_1)$ such that $f_{2, \pi(i)}$ and $f_{1, i}$ are equal up to sign for each $1 \le i \le s_1$.
\end{lemma}
\begin{proof}
    Suppose that $f_1 = \prod_{i = 1}^{s_1} f_{1, i}$ and $f_2 = \prod_{i = 1}^{s_2} f_{2, i}$ are indeed equal up to sign. If $s_1 = 0$, then it is straightforward to see that $s_2 = 0$ must also be satisfied, hence there clearly exists a trivial permutation $\pi \in \Sym(0)$ with the desired property. On the other hand, if $s_1 = 1$, then it is impossible for $s_2 \ge 2$ to be true, since this would imply that $f_2$ is a product-type expression of positive depth. In this scenario, we would have $f_1 = f_{1, 1}$, while $f_2$ would surely be distinct from both $f_{1, 1}$ and $-f_{1, 1}$, according to Theorem \ref{sum_prod_th}. Therefore, whenever $s_1 = 1$, we necessarily have $s_2 = 1$ as well, and the required permutation $\pi \in \Sym(1)$ obviously exists once again. Thus, we may now suppose without loss of generality that $s_1, s_2 \ge 2$.

    It is trivial to notice that $\Var(f_1) = \Var(f_2)$. For starters, we will show that if two variables $x_{\ell_1}, x_{\ell_2} \in \Var(f_1)$ are contained by the same operand from $f_{1, 1}, f_{1, 2}, \ldots, f_{1, s_1}$, then they are also contained by the same operand from $f_{2, 1}, f_{2, 2}, \linebreak \ldots, f_{2, s_2}$.
    The proof will be carried out by contradiction, so we may assume without loss of generality that $x_{\ell_1}, x_{\ell_2} \in \Var(f_{1, 1})$, while $x_{\ell_1} \in \Var(f_{2, 1})$ and $x_{\ell_2} \in \Var(f_{2, 2})$. Here, it is easy to see that $f_{1, 1}$ must be a sum-type expression of positive depth, which means that it can be represented as $f_{1, 1} = \sum_{1 = 1}^t g_i$, where $t \ge 2$ and $g_1, g_2, \ldots, g_t \in \mathbb{Z}(X)$ are product-type expressions with mutually disjoint sets of contained variables. Since $x_{\ell_1} \in \Var(g_k)$ clearly holds for a fixed $1 \le k \le t$, we promptly reach
    \[
        \frac{\partial f_1}{\partial x_{\ell_1}} = \frac{\partial f_{1, 1}}{\partial x_{\ell_1}} \, \prod_{i = 1, \, i \neq k}^{s_1} f_{1, i} = \frac{\partial g_k}{\partial x_{\ell_1}} \, \prod_{i = 1, \, i \neq k}^{s_1} f_{1, i} .
    \]
    Therefore, if $x_{\ell_2} \not\in \Var(g_k)$, it quickly follows that $x_{\ell_2} \not\in \Var\left(\frac{\partial f_1}{\partial x_{\ell_1}}\right)$. However, this is obviously not possible bearing in mind that
    \[
        \dfrac{\partial f_2}{\partial x_{\ell_1}} = \frac{\partial f_{2, 1}}{\partial x_{\ell_1}} \, \prod_{i = 2}^{s_2} f_{2, i} .
    \]
    Thus, we may conclude that $x_{\ell_2} \in \Var(g_k)$ necessarily holds. Due to $t \ge 2$, it follows that there must exist a variable $x_{\ell_3} \in \Var(f_{1, 1})$ such that $x_{\ell_3} \not\in \Var(g_k)$. Now, it is not difficult to apply an analogous reasoning in order to obtain that $x_{\ell_1}, x_{\ell_2} \not\in \Var\left( \frac{\partial f_1}{\partial x_{\ell_3}}\right)$. Regardless of which operand from $f_{2, 1}, f_{2, 2}, \ldots, f_{2, s_2}$ contains the variable $x_{\ell_3}$, it is impossible for $\frac{\partial f_2}{\partial x_{\ell_3}}$ to contain neither $x_{\ell_1}$ nor $x_{\ell_2}$ since these two variables are not contained by the same $f_{2, i}$-operand, thus yielding a contradiction.

    Taking everything into consideration, we have that any two variables $x_{\ell_1}, x_{\ell_2} \in \Var(f_1)$ are contained by the same $f_{1, i}$-operand if and only if they are contained by the same $f_{2, i}$-operand. From here, it quickly follows that $s_2 = s_1$ and that there exists a permutation $\pi \in \Sym(s_1)$ such that $\Var(f_{2, \pi(i)}) = \Var(f_{1, i})$ for every $1 \le i \le s_1$. In order to finalize the proof, it is enough to show that $f_{2, \pi(i)}$ and $f_{1, i}$ are necessarily equal up to sign for each $1 \le i \le s_1$. Moving on, for each fixed $1 \le k \le s_1$, we see that
    \[
        \frac{f_{2, \pi(k)}}{f_{1, k}} = \pm \prod_{i = 1, \, i \neq k}^{s_1} \frac{f_{1, i}}{f_{2, \pi(i)}} .
    \]
    Since $\Var\left(\frac{f_{2, \pi(k)}}{f_{1, k}}\right)$ is simultaneously a subset of $\Var(f_{1, k})$ and $\bigcup_{i = 1, \, i \neq k}^{s_1} \Var(f_{1, i})$, it promptly follows that $\frac{f_{2, \pi(k)}}{f_{1, k}}$ must be a constant. Corollary \ref{quotient_cor} now immediately implies that $f_{2, \pi(k)}$ and $f_{1, k}$ are equal up to sign. Therefore, $f_{2, \pi(i)}$ and $f_{1, i}$ are surely equal up to sign for any $1 \le i \le s_1$.
\end{proof}

We end the section by combining Lemma \ref{pure_product_lemma} together with Corollary \ref{quotient_cor} in order to give the proof of Theorem \ref{prod_prod_th}.

\bigskip\noindent
\emph{Proof of Theorem \ref{prod_prod_th}}.\quad
If $s_2 = s_1, \, t_2 = t_1$, and the given permutations $\pi \in \Sym(s_1)$ and $\sigma \in \Sym(t_1)$ exist, then it is clear that $\dfrac{\prod_{i=1}^{s_1} f_{1, i}}{\prod_{i=1}^{t_1} g_{1, i}}$ and $\dfrac{\prod_{i=1}^{s_2} f_{2, i}}{\prod_{i=1}^{t_2} g_{2, i}}$ are surely equal up to sign. Therefore, it is enough to only prove the converse. Suppose that the two given product-type expressions $f_1 = \dfrac{\prod_{i=1}^{s_1} f_{1, i}}{\prod_{i=1}^{t_1} g_{1, i}}$ and $f_2 = \dfrac{\prod_{i=1}^{s_2} f_{2, i}}{\prod_{i=1}^{t_2} g_{2, i}}$ are indeed equal up to sign.

To begin, observe that $\Var(f_1) = \Var(f_2)$. We will now show that if some variable $x_{\ell} \in \Var(f_1)$ is contained by one of the operands $f_{1, 1}, f_{1, 2}, \ldots, f_{1, s_1}$, then it is also certainly contained by one of the operands $f_{2, 1}, f_{2, 2}, \ldots, f_{2, s_2}$. Indeed, if this was not true, then without loss of generality, we would have $x_{\ell} \in \Var(f_{1, 1})$ and $x_{\ell} \in \Var(g_{2, 1})$. However, in this case, the fraction $\prod_{i=1}^{s_1} f_{1, i} \prod_{i=1}^{t_2} g_{2, i}$ would contain the variable $x_{\ell}$, while the fraction $\prod_{i=1}^{s_2} f_{2, i} \prod_{i=1}^{t_1} g_{1, i}$ would not, which is impossible since these two fractions are obviously equal up to sign. Therefore, each variable $x_{\ell} \in \Var(f_1)$ is contained by one of the operands $f_{1, i}, \, 1 \le i \le s_1$, if and only if it is contained by one of the operands $f_{2, i}, \, 1 \le i \le s_2$. This observation promptly implies $\bigcup_{i = 1}^{s_1} \Var(f_{1, i}) = \bigcup_{i = 1}^{s_2} \Var(f_{2, i})$, alongside $\bigcup_{i = 1}^{t_1} \Var(g_{1, i}) = \bigcup_{i = 1}^{t_2} \Var(g_{2, i})$.

Moving on, it is easy to see that $\dfrac{\prod_{i=1}^{s_1} f_{1, i}}{\prod_{i=1}^{s_2} f_{2, i}}$ and $\dfrac{\prod_{i=1}^{t_1} g_{1, i}}{\prod_{i=1}^{t_2} g_{2, i}}$ are equal up to sign. Moreover, we know that the sets $\Var\left( \dfrac{\prod_{i=1}^{s_1} f_{1, i}}{\prod_{i=1}^{s_2} f_{2, i}} \right)$ and $\Var\left( \dfrac{\prod_{i=1}^{t_1} g_{1, i}}{\prod_{i=1}^{t_2} g_{2, i}} \right)$  are necessarily disjoint, thus implying that $\dfrac{\prod_{i=1}^{s_1} f_{1, i}}{\prod_{i=1}^{s_2} f_{2, i}}$ must be a constant. If $s_1 = 1$, then $\prod_{i=1}^{s_1} f_{1, i} = f_{1, 1}$ is obviously a sum-type expression, while for $s_1 \ge 2$, we have that $\prod_{i=1}^{s_1} f_{1, i}$ represents a product-type expression. The same conclusion can be made for $\prod_{i=1}^{s_2} f_{2, i}$, which means that both of these fractions are certainly a sum-type or product-type expression. By virtue of Corollary \ref{quotient_cor}, we may conclude that $\prod_{i=1}^{s_1} f_{1, i}$ and $\prod_{i=1}^{s_2} f_{2, i}$ are equal up to sign, which means that $\prod_{i=1}^{t_1} g_{1, i}$ and $\prod_{i=1}^{t_2} g_{2, i}$ are as well. As a direct consequence of Lemma \ref{pure_product_lemma}, we obtain that $s_2 = s_1, \, t_2 = t_1$, and that there exist permutations $\pi \in \Sym(s_1)$ and $\sigma \in \Sym(t_1)$ such that $f_{2, \pi(i)}$ and $f_{1, i}$ are equal up to sign for each $1 \le i \le s_1$, while $g_{2, \sigma(i)}$ and $g_{1, i}$ are equal up to sign for any $1 \le i \le t_1$. \hfill\qed

\section{Algorithm overview}\label{sc_algorithm}

The main idea behind the algorithm is to iteratively compute the number of sum-type and product-type expressions on $k$ fixed variables for all positive integers $k$ up to $n \in \mathbb{N}$. We denote the number of sum-type expressions on $k$ fixed variables by $S_k$, the number of product-type expressions on $k$ fixed variables by $P_k$ and the number of expressions of either type by $A_k$. In order to determine these values, we will also need to have the binomial coefficients $\binom{k}{m}$ computed for all $k, m \in \mathbb{N}_0, 0 \leq m \leq k \leq n$. We will show that it is sufficient to store $O(n)$ of these binomial coefficients in memory at any given time. The algorithm is a direct implementation of all the propositions given throughout this section. To begin, we derive the following result.

\begin{proposition} \label{num_sum_exprs}
    The number of sum-type expressions on $k \geq 2$ fixed variables is given by
    \begin{equation}\label{algaux_1}
        S_k = \sum_{j=1}^{k-1} \binom{k-1}{j-1} P_j A_{k-j}.
    \end{equation}
\end{proposition}
\begin{proof}
    Without loss of generality, let $f$ be a sum-type expression on the variables $\{x_1, x_2, \ldots, x_k\}$, where $k \ge 2$. It is obvious that $f$ must be a sum-type expression of positive depth. Due to Theorem \ref{sum_sum_th}, it immediately follows that $f$ can be uniquely decomposed into $s \ge 2$ summands $f_1, f_2, \ldots, f_s$ that represent product-type expressions with mutually disjoint sets of contained variables. Clearly, we may reorder these terms so that $x_1 \in \Var(f_1)$. Since there are at least two terms, $f_1$ may contain anywhere between one and $k-1$ variables, inclusive. Let $j$ be this number of variables. There are $\binom{k-1}{j-1}$ ways to choose $j-1$ variables out of the remaining $k-1$ variables for $f_1$. The first term is a product-type expression, so there are $P_j$ inequivalent ways to construct it using the $j$ chosen variables. The sum $\sum_{i=2}^s f_i$ by itself is either a sum-type expression of positive depth (if $s \ge 3$) or a product-type expression (if $s = 2$). In effect, due to Theorem \ref{sum_prod_th}, we can simply use the value $A_{k-j}$ to give us the required number of ways to construct the required expression out of the reminaing $k-j$ variables. By multiplying $\binom{k-1}{j-1}$, $P_j$ and $A_{k-j}$ and summing up over all $j, \, 1 \leq j \leq k-1$, we obtain Equation \eqref{algaux_1}.
\end{proof}

Before we present an analogous formula for the number of product-type expressions, we shall define two more auxiliary types of expressions:\ $\Pi_2$-type expressions and $\Pi_1$-type expressions, and then give a formula for counting them. Unlike other types of expressions, we will consider two expressions of these two types to be equivalent provided they are equal up to sign, rather than just equal.

\begin{definition}
    An expression is a \emph{$\Pi_2$-type expression} if it is a product of two or more sum-type expressions with mutually disjoint sets of contained variables.
\end{definition}

\begin{definition}
    An expression is a \emph{$\Pi_1$-type expression} if it is either a $\Pi_2$-type expression or a sum-type expression.
\end{definition}

We will denote the number of inequivalent $\Pi_2$-type expressions on $k$ fixed variables by $Q_k$ and the number of inequivalent $\Pi_1$-type expressions on $k$ fixed variables by $R_k$. The number of inequivalent sum-type expressions on $k$ fixed variables, up to sign, is precisely $\frac{S_k}{2}$, because they can obviously be split into additively complementary pairs. Therefore, by Theorem \ref{sum_prod_th}, for any $k \geq 2$ we have $R_k = Q_k + \frac{S_k}{2}$. Moving on, we will rely on the next proposition to count the number of $\Pi_2$-type expressions on $k$ fixed variables.

\begin{proposition}
    The number of inequivalent $\Pi_2$-type expressions on $k \ge 2$ fixed variables is given by
    \begin{equation}\label{algaux_2}
        Q_k = \sum_{j=1}^{k-1} \binom{k-1}{j-1} \frac{S_j}{2} R_{k-j}.
    \end{equation}
\end{proposition}
\begin{proof}
    By virtue of Theorem \ref{prod_prod_th}, we observe that any $\Pi_2$-type expression is surely a product-type expression of positive depth. Analogously to the proof of Proposition \ref{num_sum_exprs}, we may assume that the first factor is a sum-type expression, while the remaining factors necessarily form a $\Pi_1$-type expression. Equation~\eqref{algaux_2} follows immediately from here. Note that $S_j$ should be divided by two in order to count only inequivalent sum-type expressions up to sign. While counting the $\Pi_1$-type expressions, the sign is already ignored.
\end{proof}

Next, we present a formula for computing the number of product-type expressions on $k$ fixed variables.

\begin{proposition}
    The number of product-type expressions on $k \ge 2$ fixed variables is given by
    \begin{equation}\label{algaux_3}
        P_k = 2\left(Q_k + \sum_{j=1}^{k-1} \binom{k}{j} R_j R_{k-j}\right).
    \end{equation}
\end{proposition}
\begin{proof}
    We will count the number of inequivalent product-type expressions up to sign and then just multiply this result by two. It is easy to notice that any product-type expression on $k \ge 2$ fixed variables is surely of positive depth. Therefore, due to Theorem \ref{prod_prod_th}, it follows that such product-type expressions inequivalent up to sign can be uniquely decomposed as $\dfrac{\prod_{i=1}^{s} f_{i}}{\prod_{i=1}^{t} g_{i}}$, where $s \ge 1, t \ge 0, s + t \ge 2$, and $f_1, \ldots, f_s, g_1, \ldots, g_t$ are sum-type expressions with mutually disjoint sets of contained variables. For $t = 0$, it is clear that the number of desired expressions equals $Q_k$. Now, suppose that $t \ge 1$ and let $j$ be the number of variables appearing in the $f_1, f_2, \ldots, f_s$ factors. Observe that in this case, $j$ can be any integer between one and $k-1$, inclusive. Furthermore, there are $\binom{k}{j}$ ways to select these $j$ variables and then $R_j$ and $R_{k-j}$ ways to construct the $\Pi_1$-type expressions $\prod_{i=1}^s f_i$ and $\prod_{i=1}^t g_i$, respectively. Equation~\eqref{algaux_3} follows directly from here.
\end{proof}

Finally, as a direct consequence of Theorem \ref{sum_prod_th}, we also reach the following result.

\begin{proposition}
    The number of sum-type or product-type expressions on $k \geq 2$ fixed variables is given by
    \[
        A_k = S_k + P_k.
    \]
\end{proposition}

It is trivial to obtain the values $S_1 = 2, \, Q_1 = 1, \, R_1 = 1, \, P_1 = 2, \, A_1 = 2$. Now, let us find the order in which the values $S_k, Q_k, P_k, R_k, A_k$ can be computed for all $k \ge 2$. For any given $k \geq 2$, we can compute $S_k, Q_k$ using the values $A_j, P_j, R_j, S_j$ for $1 \le j < k$, and the binomial coefficients of the form $\binom{k-1}{j}$ for all the $0 \leq j \leq k-1$. This sequence of binomial coefficients is also known as the $(k-1)$-th row of the Pascal triangle (see, for example, \cite{stanley2011}). Subsequently, we can compute $R_k = Q_k + \frac{S_k}{2}$. Afterwards, we may determine $P_k$, as Equation~\eqref{algaux_3} relies only on $Q_k$, the values $R_j$ for $1 \le j < k$, and the $k$-th row of the Pascal triangle. Finally, we determine $A_k$. Note that, in order to compute all the terms $P_k, Q_k, R_k, S_k, A_k$, we only need to store the $(k-1)$-th and $k$-th row of the Pascal triangle in memory. We can compute the $(k+1)$-th row from the $k$-th row by implementing the well-known identity
\[
    \binom{k+1}{j} = \binom{k}{j} + \binom{k}{j-1}.
\]
Before we do this, we can discard the $(k-1)$-th row from memory as it is no longer needed. In order to find the values $A_k$ for all $1 \leq k \leq n$, at any moment we only need to store the values $P_j, Q_j, R_j, S_j, A_j$ for all the $1 \le j \le n$, two rows of the Pascal triangle that both contain at most $n + 1$ elements and a constant number of auxilliary variables.

Therefore, the memory complexity of the disclosed algorithm is indeed $\Theta(n)$, measured as the number of stored arbitrary precision integers, while its time complexity is obviously $\Theta(n^2)$, measured as the number of arithmetic operations on arbitrary precision integers. An example of a C++ implementation of the algorithm as a function template can be found in Appendix \ref{sc_code}. For an overview of C++ templates, please refer to \cite{vandevoorde2017}.

\section{Some open problems}\label{sc_conclusion}

Throughout the paper, we have proved various properties concerning the equivalence of sum-type and product-type expressions. These results lead up to the construction of a $\Theta(n^2)$ algorithm for computing the number of inequivalent arithmetic expressions on $k$ distinct variables containing the standard four arithmetic operations alongside additive inversion, for every $1 \le k \le n$. Such an algorithm represents an efficient complete solution to the problem considered by Radcliffe \cite{radcliffe2012}. We end the paper by disclosing some open problems directly connected to the aforementioned question.

For starters, it is natural to ask whether a similar approach could be applied to construct a $\Theta(n^2)$ algorithm for computing the number of inequivalent arithmetic expressions on $n$ distinct variables when the only allowed operations are addition, subtraction, multiplication and division, without additive inversion. This problem was actually considered by Du \cite{du2008} and although its formulation is quite similar, it seems that the algorithm shown in Section \ref{sc_algorithm} cannot directly be modified to provide the desired solution. For this reason, we choose to leave the following related question.

\begin{question}
    Can we construct a $\Theta(n^2)$ algorithm for computing the number of inequivalent arithmetic expressions on $k$ distinct variables, for each $1 \le k \le n$, when the only allowed arithmetic operations are the standard four binary operations, without additive inversion?
\end{question}

Moving on, it makes sense to consider the similar problem of computing the number of inequivalent arithmetic expressions on a given number of variables when it is not guaranteed that all the variables are mutually distinct. This immediately leads us to the next two open problems.

\begin{problem}
    Find an efficient algorithm for computing the number of inequivalent arithmetic expressions that can be represented by an expression tree such that:
    \begin{enumerate}[label=\textbf{(\roman*)}]
        \item the allowed operations are the standard four arithmetic operations, i.e., addition, subtraction, multiplication and division, alongside additive inversion;
        \item there are $\sum_{i = 1}^{n} \alpha_i$ leaves, each containing one of the $n \in \mathbb{N}$ formal variables $\{ x_1, x_2, \ldots, x_n\}$, so that $x_i$ appears $\alpha_i \in \mathbb{N}$ times, for any $1 \le i \le n$.
    \end{enumerate}
\end{problem}

\begin{problem}
    Find an efficient algorithm for computing the number of inequivalent arithmetic expressions that can be represented by an expression tree such that:
    \begin{enumerate}[label=\textbf{(\roman*)}]
        \item the allowed operations are the standard four arithmetic operations, i.e., addition, subtraction, multiplication and division, without additive inversion;
        \item there are $\sum_{i = 1}^{n} \alpha_i$ leaves, each containing one of the $n \in \mathbb{N}$ formal variables $\{ x_1, x_2, \ldots, x_n\}$, so that $x_i$ appears $\alpha_i \in \mathbb{N}$ times, for any $1 \le i \le n$.
    \end{enumerate}
\end{problem}

Finally, we shall mention the question of finding the number of expressions inequivalent up to permutation of variables. More precisely, for each bijection $\varphi \colon X \to X$, let $\mathcal{F}_{\varphi} \colon \mathbb{Z}(X) \to \mathbb{Z}(X)$ denote the mapping that transforms each fraction by replacing the formal variable $x_i$ with $\varphi(x_i)$, for every $i \in \mathbb{N}$. It is now convenient to define two fractions $f_1, f_2 \in \mathbb{Z}(X)$ to be \emph{symmetrically equivalent} if there exists a bijection $\varphi \colon X \to X$ such that $f_2 = \mathcal{F}_{\varphi}(f_1)$. With this in mind, we disclose the following two open problems.

\begin{problem}\label{sym_prob_1}
    Find an efficient algorithm for computing the number of symmetrically inequivalent arithmetic expressions on $n \in \mathbb{N}$ distinct variables, when the only allowed arithmetic operations are addition, subtraction, multiplication, division and additive inversion.
\end{problem}

\begin{problem}\label{sym_prob_2}
    Find an efficient algorithm for computing the number of symmetrically inequivalent arithmetic expressions on $n \in \mathbb{N}$ distinct variables, when the only allowed arithmetic operations are addition, subtraction, multiplication and division, without additive inversion.
\end{problem}

\begin{remark}
    Of course, it is also possible to consider the closely related variants of Problems \ref{sym_prob_1} and \ref{sym_prob_2} when the formal variables are not guaranteed to be mutually distinct.
\end{remark}

\section*{Acknowledgements}
The authors express their gratitude to Nino Bašić for his useful comments and remarks. Ivan Damnjanović is supported by the Science Fund of the Republic of Serbia, grant \#6767, Lazy walk counts and spectral radius of threshold graphs --- LZWK.

\section*{Conflict of interest}

The authors declare that they have no conflict of interest.

\appendix
\section{Implementation in C++}\label{sc_code}

\begin{lstlisting}[language = C++, frame = trBL, escapeinside={(*@}{@*)}, aboveskip=10pt, belowskip=10pt, numbers=left, rulecolor=\color{black}]
template <class Number>
std::vector<Number> unaryMinusAllowed(size_t n) {
  std::vector<Number> A(n + 1), S(n + 1), Q(n + 1),
    R(n + 1), P(n + 1);
  A[1] = S[1] = P[1] = 2;
  Q[1] = R[1] = 1;

  // Binomial coefficients (two rows of Pascal's triangle)
  std::vector<Number> Bk = {1, 2, 1}, Bkm1 = {1, 1};

  for (size_t k = 2; k <= n; k++) {
    // Number of sum expressions
    Number total = 0;
    for (size_t j = 1; j <= k-1; j++)
      total += P[j] * A[k - j] * Bkm1[j - 1];
    S[k] = total;
    // Number of Pi-2 type expressions
    total = 0;
    for (size_t j = 1; j <= k-1; j++)
      total += S[j] * R[k - j] * Bkm1[j - 1];
    Q[k] = total / 2;
    // Number of product expressions
    total = Q[k];
    for (size_t j = 1; j <= k-1; j++)
      total += R[j] * R[k - j] * Bk[j];
    P[k] = total * 2;
    // Number of Pi-1 and any type expressions
    R[k] = Q[k] + S[k] / 2;
    A[k] = S[k] + P[k];
    // Update binomial coefficients rows
    Bkm1 = std::move(Bk);
    Bk.resize(k + 2);
    for (size_t j = 1; j <= k; j++)
      Bk[j] = Bkm1[j] + Bkm1[j - 1];
    Bk[0] = Bk[k + 1] = 1;
  }

  return A;
}
\end{lstlisting}


\begin{thebibliography}{99}

\bibitem{du2008} Z.H.\ Du, Sequence A140606 in {\em The On-Line Encyclopedia of Integer Sequences}, 2008, \url{https://oeis.org/A140606}.

\bibitem{dufo2004} D.S.\ Dummit and R.M.\ Foote, \emph{Abstract Algebra}, John Wiley \& Sons, Inc., 3\textsuperscript{rd} edition, 2004.

\bibitem{gopal2010} A.\ Gopal, \emph{Magnifying Data Structures}, PHI Learning Private Limited, New Delhi, 2010.

\bibitem{grune2012} D.\ Grune, K.\ van Reeuwijk, H.E.\ Bal, C.J.H.\ Jacobs and K.\ Langendoen, \emph{Modern Compiler Design}, Springer New York, NY, 2\textsuperscript{nd} edition, 2012, \linebreak \doi{10.1007/978-1-4614-4699-6}.

\bibitem{lang2002} S.\ Lang, \emph{Algebra}, volume 211 of \emph{Graduate Texts in Mathematics}, Springer New York, NY, 3\textsuperscript{rd} edition, 2002, \doi{10.1007/978-1-4613-0041-0}.

\bibitem{pinter2009} C.C.\ Pinter, \emph{A Book of Abstract Algebra}, Dover Publications, Inc., Mineola, New York, 2\textsuperscript{nd} edition, 2009.

\bibitem{preiss1999} B.R.\ Preiss, \emph{Data Structures and Algorithms with Object-Oriented Design Patterns in Java}, John Wiley \& Sons, Inc., 1999.

\bibitem{radcliffe2012} D.\ Radcliffe, Sequence A182173 in {\em The On-Line Encyclopedia of Integer Sequences}, 2012, \url{https://oeis.org/A182173}.

\bibitem{stanley2011} R.P.\ Stanley, \emph{Enumerative Combinatorics:\ Volume 1}, Cambridge University Press, 2\textsuperscript{nd} edition, 2011, \doi{10.1017/CBO9781139058520}.

\bibitem{vandevoorde2017} D.\ Vandevoorde, N.\ Josuttis and D.\ Gregor, \emph{C++ Templates:\ The Complete Guide}, Addison-Wesley Professional, 2\textsuperscript{nd} edition, 2017.

\bibitem{zhang2018_a} Z.\ Zhang, 2018, \url{https://zhuanlan.zhihu.com/p/34058293}.

\bibitem{zhang2018_b} Z.\ Zhang, 2018, \url{https://zhuanlan.zhihu.com/p/34261468}.

\end{thebibliography}
\end{document}